\documentstyle[aas2pp4]{article}

\lefthead{Simkin, Sadler, Sault, Tingay and Callcut}
\righthead{Pictor A}

\begin{document}
\slugcomment{Submitted to the Ap.J., October 6, 1998, Accepted March 17, 1999}

\title{Pictor A (PKS 0518$-$45) - From Nucleus to Lobes}

\author{S. M. Simkin
\altaffilmark{1,2,3}} 
\affil{Michigan State University, East Lansing, MI 48824-1116}
\authoremail{simkin@grus.pa.msu.edu}

\author{E. M. Sadler
\altaffilmark{1,3}} 
\affil{School of Physics, University of Sydney, NSW 2006, Australia}

\author{R. Sault
\altaffilmark{3}} 
\affil{Australia Telescope National Facility, CSIRO, Epping, NSW 1710
 Australia}

\author{S. J. Tingay
\altaffilmark{}} 
\affil{California Institute of Technology, Jet Propulsion Laboratory (MS238-332), 4800 Oak Grove Drive, 
Pasadena, CA 91109 }

\and

\author{J. Callcut\altaffilmark{4}}
\affil{Michigan State University, East Lansing, MI 48824-1116}

\altaffiltext{1}{Visiting Astronomer, Cerro Tololo Inter-American Observatory. 
CTIO is operated by AURA, Inc.\ under contract to the National Science
Foundation.} 
\altaffiltext{2}{Visiting Astronomer, Space Telescope Science
Institute, STScI is operated by AURA, Inc.\ under contract to NASA}
\altaffiltext{3}{Visiting Astronomer, The Austraila Telescope, funded
by the Commonwealth of Australia for operation as a National Facility
managed by CSI }
\altaffiltext{4}{NSF REU summer student, 1992}

\begin{abstract}

We present a variety of new imaging and kinematic data for the
double--lobed radio galaxy Pictor A. The new optical data include HST
continuum and [OIII], emission--line images (at a resolution of 25 --
100 mas) and ground-based imaging and spectroscopy (at a resolution of
$\sim 1.5 \arcsec$). The ground--based images show H$\alpha$ filaments
and loops which extend to the N and W of the optical core.  The radio
continuum data include 3\,cm Australia Telescope images of the core, at
a resolution comparable to that of the optical, ground--based images,
and  a VLBI image of a jet in the compact core (at a resolution of 2 --
25 mas), which seems to align with a continuum ``jet'' found in the HST
images.  The core radio jet, the HST optical continuum ``jet'', and the
NW H$\alpha$ filaments all appear to be aligned with the extended, low
luminosity radio continuum bridge which \cite{Pr97} have traced out to
the optical--synchrotron hot--spot in the NW lobe of this object. The
H$\alpha$ filaments which appear to lie in the path of this trajectory are
associated with a disrupted velocity field in the extended ionized
gas.  These filaments (as well as the extended [OIII] emission found at
mas scales in the HST images) may have been pushed to the N, out
of the path of the jet.  The ground--based spectra  which cover this
trajectory also yield line ratios for the ionized gas which have
anomalously low [NII] (6564), suggesting either a complex, clumpy
structure in the gas with a higher cloud-covering factor at larger
radii and with denser clouds than is found in the nuclear regions of
most NLRG and Seyfert 2 galaxies, or some other, unmodeled, mechanism
for the emergent spectrum from this region. The H$\alpha$
emission--line filaments to the N appear to be associated with a  3\,cm
radio continuum knot which lies in a {\it gap} in the filaments $\sim 4
\arcsec$ from the nucleus.  Altogether, the data in this paper provide
good circumstantial evidence for non--disruptive redirection of a
radio jet by interstellar gas clouds in the host galaxy.

\end{abstract}


\section{Introduction}

\subsection{Background}

Pictor A is the closest, double--lobed radio source with a
well--defined jet and an active, optical--synchrotron hot--spot
(\cite{MR87,Th95}).  Recent, high--resolution, multi--frequency VLA
radio images find a ``bridge'' or channel of low--level radio emission
which links this hot--spot to the nuclear region  (\cite{Pr97}).
 It is the seventh brightest radio source in the sky at 408\,MHz and
the most powerful radio galaxy in the redshift range $0<z<0.04$.  It is
also the nearest broad--lined radio galaxy with a nuclear, optical
spectrum which has broad, permitted emission lines (H, He), and
relatively narrow optical forbidden emission lines ([OI], [SII],
[OIII], [NII]) superimposed on a featureless continuum.  The broad
permitted lines have FW0I$\,>\,14,000$\,km\,s$^{-1}$ (from our CTIO
measurements reported in this paper), but are reported to be as high as
23,000\,km\,s$^{-1}$ (\cite{Fp85,Su95,Ha96} and references therein),
with a highly excited nuclear spectrum. It has a compact radio core
(\cite{Ch77}).  Its redshift is $z = 0.035$ ($v = 10,500$ km$s^{-1}$ ),
giving a distance, D = 110 Mpc, if H$_0$ = 100 km$s^{-1}$ $Mpc^{-1}$
and q$_0$ = 0. All distance-dependent values in this paper have been
scaled to H$_0$=100 with the appropriate scaling factor,
h={10}$^{-2}$H$_0$. (With this value of $H_0$, the physical scale for
Pictor A is roughly 530\,pc$({\arcsec})^{-1}$$h^{-1}$.

\subsection{Extended Optical Structure (EELG)}

This object has been identified as one of a class of double--lobed
radio galaxies which contain extensive regions of non-nuclear, ionized
gas (cf. \cite{Ca84,Ro87,Baa89,Bab89} and references therein.) This
extended, ionized gas (EELG) is known to have a peculiar, apparently
non--rotational, velocity field (\cite{Ta89}). The extended
emission--line gas has been imaged at low resolution and with low
signal--to--noise in the redshifted emission lines of both
H$\alpha$--[NII] and [OIII]--$\lambda$5007 as well as continuum
emission (\cite{Ha87,FB91}.)

There are now several examples of nearby Seyfert galaxies with
well--studied, non--nuclear, extended emission line regions
(\cite{Al98} and references therein). There are also several studies of
the EELG associated with extended radio emission for intermediate
redshift radio galaxies (\cite{Vi98,Al98} and references therein).
There are, however, very few nearby, powerful, double--lobed radio
galaxies (FRII objects) and thus extensive studies of this type of
object are lacking.

\subsection{Motivation for the Present Study}

FRII galaxies are important because they represent one extreme edge of
the phenomenological envelope which defines ``active galaxies.'' They
are rare because the radio--loud fraction of the AGN population is
small. Thus the opportunity to study such an object with high spatial
resolution is also rare.  Because it is both nearby and an extreme
example of a powerful, almost quasar--like, radio galaxy, we regard
Pictor A as an excellent, natural laboratory for identifying those
features which differentiate radio--loud AGN from their radio--quiet
counterparts and for studying the structure of the beamed radio plasma
which produces the strong radio hotspots seen in FR\,II sources.  In
this paper we combine a wide variety of different observations in an
effort to elucidate the relationship between the jet, lobes, and
peculiar optical structure and kinematics in this galaxy. We believe
that this type of approach, where diverse observations are analyzed and
critically examined in one body of work, can lead to a much better
understanding of the interrelationships between these different
characteristics than a more piecemeal approach.

We began our study of this object as part of a program to obtain
detailed measurements of the velocity fields in FR\,II radio galaxies
with extended ionized gas. Although initial observations of several
such objects suggested a simple relationship between internal,
rotational kinematics and the axis of radio emission (\cite{Si79}, it
became clear that the situation was much more complex (cf. the
discussions in \cite{He85} and \cite{Bab89}) and warranted a program of
extensive mapping of the velocity fields in several well--resolved
objects.  Our new study, which included narrowband, optical imaging
(section 2.1) and spectroscopy (section 2.3), evolved into a much more
extensive one which included HST imaging (section 3) and, finally,
radio imaging (sections 2.2 and 4.4).

For the sake of brevity, we have attempted to limit our discussion of
observational reduction techniques in this paper (although they are
critical to understanding the limitations of the present data),
referring to other published (or soon to be published) sources where
possible. The one exception to this limited discussion of the data
analysis is our treatment of the HST image reduction in Section 3.3.
This analysis is unique and cannot be replicated unless and until the
HST is again equipped with a camera and filters which can be used for
redshifted, emission--line imaging of this object.

In the final analysis, the goal of this study is to identify which
measurable, physical parameters are critical for understanding the
radio galaxy phenomenon. Once identified, these can be incorporated
into more focused observational surveys covering a more complete
sample, which will allow a more general understanding of the physical
basis for the type of extreme radio activity seen in classical FR\,II
radio galaxies.

\section{Ground-Based Observations (CTIO and ATCA)}

\subsection{Direct Optical Images (Continuum and H$\alpha$)}

Direct images of Pictor A were taken in February 1987 with the CTIO
1.5m, 800$\times$800 TI CCD and $f/7.5$ cassegrain secondary.  The CCD
scale was 0.27 $''${pixel}$^{-1}$. The telescope PSF showed evidence of
coma (due to mis-alignment of the secondary) but the image cores were
well-defined.

Imaging was done through two interference filters, $\lambda$6560
($\Delta\lambda$110\AA), covering the continuum, and  $\lambda$6826
($\Delta\lambda$78\AA), covering redshifted [SII]--H$\alpha$--[NII].
We used the narrow band $\lambda$6560 filter to obtain continuum
observations because it provides a better match for the $\lambda$6826
filter in terms of PSF formation and exposure effects than a
broad--band filter.  The $\lambda$6560 filter does, however, include
the redshifted [OI]\/$\lambda$6364 emission line at 6587\AA, which
appears to extend beyond the nucleus. However, this should introduce
minimal distortion of the H$\alpha$ image (if any) since even in the
nucleus the strength of the [OI]\/$\lambda$6364 line is less than 7.5\%
of that from H$\alpha$ (table~\ref{tbl-3}). The imaging observations
are listed in table~\ref{tbl-1}.

\placetable{tbl-1}

The images were reduced using the ``IRAF'' package in the standard way
(correcting for detector response by removing biases, dark current, and
normalizing with ``dome'' flats). The reductions for the entire set of
CTIO images are described in more detail in \cite{Gr99}. The H$\alpha$
images were corrected for continuum  flux by  normalizing the stellar
images in the $\lambda$6560 filter to the same relative flux as those
in the $\lambda$6826 filter and then subtracting. The continuum and
corrected H$\alpha$ images are shown in Figure\,1a (continuum
isophotes) and 1b (H$\alpha$ grayscale). The H$\alpha$ image shows a
wealth of arc--like, detailed structure in the ionized gas throughout
the inner 3$''$ to 8$''$ (1.6 to 4.5 $h^{-1}$\,kpc) of the nucleus.
Although several nearby radio galaxy hosts show organized, arc-like
structure in their extended ionized gas (cf. PKS0634-20 and 3C227 in
\cite{Ha87,Ba88} and \cite{Pr93}), none show such a large, complete
loop such as this.

\placefigure{fig1}

Three points are noteworthy: 

\begin{enumerate}

\item {\it Offset in the Continuum Image:\,\,} The $\lambda$6560\,\AA \,\,continuum
image (Figure\,1a) has a much steeper profile toward the S than
the N. As we will see in section 2.2, this offset may well be related
to an important new feature which we have identified in the 3\,cm
radio image.

\item {\it H$\alpha$ Pseudo--Ring to the N:\,\,} The H$\alpha$ image
(Figure\,1b)  shows a loop of emission reaching 4$\arcsec$ to
the N with a gap at PA --15$^{o}$. This was seen in the much lower
resolution image of  \cite{Ha87} as two low--luminosity ``horns'' on
either side of the central image. Here it is clearly reminiscent  of the
types of rings, or ``pseudo'' rings, seen in some barred spirals and
Seyfert galaxies (\cite{Su80}.) Again, as we will see in section 2.2, it
may be intimately related to an unusual 3\,cm radio continuum feature.

\item {\it H$\alpha$ Bridge to the WNW:\,\,} There is also an open arc
or bridge extending to the W in the H$\alpha$ image
(Figure\,1b).  This feature (at PA 283$^{o}$) ends  in a knot
at $\sim$\,7.8$\arcsec$ from the nucleus. (This seems to have first
been seen in spectra by Baldwin and Compusano as reported in
\cite{Ca84} and, with hindsight, appears in the lower resolution in the
image by \cite{Ha87}.)

\end{enumerate}

\placefigure{fig2}

The relationship of these optical features to the radio continuum
emission is shown in figures 2 and 3. In Figure\,2, the 20\,cm
VLA, radio image of \cite{Pr97} (resolution 7.5$\arcsec$) is displayed
with a scale which shows the bridge linking the nucleus to the NW radio
hot--spot. The inset in  Figure\,2 shows isophotes from the
H$\alpha$ image in Figure\,1b with the direction of the NW
hot--spot (at PA 280--281$^{o}$) and the counter direction of the SE
hot--spot (at PA 288--289$^{o}$) shown by the arrows. The H$\alpha$
extension  at PA 283$^{o}$ as well as the 20\,cm radio bridge are both
{\it nearly} aligned with the  direction of the hotspots.

The relationship between the N {H$\alpha$ pseudo--ring and the radio
continuum is discussed below:

\subsection{Radio imaging (ATCA)}

Pictor A has no discernible optical continuum jet on arcsecond scales
(\cite{FB91}), and although the 20\, cm observations of \cite{Pr97}, 
plotted in  Figure\,2, show  a kpc--scale jet pointing to the
nucleus from the NW hot--spot, they do not reveal a small--scale jet.
However, the H$\alpha$ feature shown in  Figure\,2 and the
continuum feature seen in our HST Cycle 1 observations (section 3) are
so well aligned with the 130\,kpc--distant, NW hotspot that we made
three 12--hour 3\,cm synthesis observations with the Australia Telescope
Compact Array (ATCA) to search for radio continuum features which might
correlate with the optical features we found in the CTIO and HST data.

\subsubsection{Observations}

The ATCA is an east-west, 6-antenna,
interferometer array located in south-eastern Australia (\cite{AT92}).
Three array configurations were used (the so-called 6A, 6C and 6D
configurations), which give good Fourier plane coverage out to 6 km.
The observations were on 1993 Jan 10, 1993 Feb 11 and 1993 Sep 20.  A
fourth configuration was rendered unusable by an error in the on-line
system.  The flux scale was set using PKSB1934-638 (\cite{Re94}), whereas
PKSB0534-340 was used as the phase calibrator.  The technique of
multi-frequency synthesis (\cite{Co90}, \cite{Sa94}) was used to improve
the Fourier plane coverage beyond that which would normally be achieved
with the available configurations.  Multi-frequency synthesis relies on
measuring different Fourier spacings by varying the observing
frequency.  In each observing run, we alternated between two setups
every 5 min, with each setup observing two different frequencies
simultaneously (i.e.  in each run, we observed four distinct
frequencies).  The observing frequencies were in the range to 8.24 to
9.11 GHz, with the weighted mean being 8.62 GHz.

\subsubsection{Reductions and Analysis}

The data were flagged and calibrated in the Miriad system (\cite{Sa95}),
with each frequency calibrated separately.  Images were formed,
deconvolved and self-calibrated in a fashion similar to that used by
\cite{Sa94}.  This procedure determines and corrects for the effects of
non-zero spectral indices in multi-frequency synthesis. The shortest
spacing used in the imaging was 153 m, which resolves out the very
extended emission seen by \cite{Pr97}.

The resultant images have moderate dynamic range in the region of the
core (1:2000), which is worse than the thermal limit by a factor of 6.
The limit in the reconstruction is believed to be related to the size
of the object. The north-east hot-spot occurs well out in the primary
beam (the north-east hot-spot is 250 arcmin from the core, whereas the
half-width at half-maximum of the primary beam response is about 176
arcmin). Deviations from the primary beam response being circularly
symmetric this far out, as well as possible pointing problems, have
resulted in the imaging of the north-east hot-spots being quite poor
(Figure\,3.)

We searched for but found no evidence for variation of the flux density
of the core (0.92 Jy at 8.62 GHz) over the 8-month period of our
observations.

\placefigure{fig3}

\subsubsection{Radio Images}

The resulting 3\,cm radio map is shown in Figure\,3. The inset
to Figure\,3 shows the core of this image plotted on top of the
gray--scale H$\alpha$ image shown in Figure\,1b. The
radio--optical images have been registered so that the peak optical
(nuclear) brightness corresponds to the measured peak in the radio
core. This introduces an offset of $\alpha$=0.4$\arcsec$ and
$\delta$=\/-1.4$\arcsec$ (optical -- radio).  This is well within the
accuracy of the HST--GSC 1.0 optical position (\cite{La90} ,
\cite{Je90}, \cite{Ru90}).   With this offset, we note a lobe (located
at PA = -15$^{o}$ and 4$\arcsec$ N of the core) {\it the peak of which
is coincident with the gap in the H$\alpha$ loop at that position}
(Figure\,3--inset). Although this lobe is weak (it is weaker
than some of the artifacts in the north-east and south-west hot-spots),
its peak is ten times the rms in the vicinity of the core, it appears
in images made from the three configurations individually, and it is
robust to various self-calibration models. Hence we believe this lobe
is not an artifact.

\subsection{CTIO spectra}

Nine long--slit spectra were taken during four consecutive nights in
February 1987 with the CTIO 4m spectrograph and the TI CCD. Two
additional spectra were taken with the same instrument in January 1988.
All exposures were bracketed with spectra of a HeArNe arc lamp taken
with  the spectrograph in the same PA as the galaxy spectrum and the
telescope at the sky positions corresponding to the start and end
positions for the intervening galaxy spectrum.  The slit positions and
observing details  are shown in table~\ref{tbl-2}. In all cases the
scale along the slit was 0.73 $''$/pixel; The slit width was 180
$\mu$m (1.2 $''$). All galaxy exposures were 2000 seconds.

\placetable{tbl-2}

\subsubsection{Calibration}

The spectra were flat--fielded, calibrated in wavelength using the arc
spectra, and ``scrunched'' to remove S-distortion.  The internal
consistency in the  fits to the calibration data was $\pm 0.37$\AA
for the blue-green spectra and $\pm 0.30$\AA  for the red spectra.
The final dispersions were 1.49\,\AA\,{pixel}$^{-1}$ and
1.21\,\AA\,{pixel,}$^{-1}$ respectively. The calibrated spectra were
used to measure wavelength positions for the emission lines H$\alpha$,
$\beta$, [O\,I], [O\,III], [N\,II], and [S\,II].  These were then used
to calculate absolute velocities at each position (stepping 0.73$''$)
from the nucleus where the observed emission lines were at least 3
times the estimated ``noise'' in the adjacent continuum. The internal
error estimates noted above yield a formal error for the velocity
determinations of 14 km$s^{-1}$ for the red emission lines and 21
km$s^{-1}$ for the blue.  Since these do not include systematic
effects, one might expect these estimates to define a lower limit to
the velocity measurements.  However, the average nuclear velocity from
all of the measurements is 10548 $\pm 21$ km$s^{-1}$ (section 2.3.2),
suggesting that the CTIO spectrograph was quite stable over the 13
month period when the observations were taken. (We note that this {\it
internal} accuracy does not measure any {\it systematic} shifts in the
velocity zero points.)

Although we observed flux calibration standards for all of these
spectra, it is clear from examination and comparison of these
calibration spectra from night to night that the nights were
non-photometric, with flux variations of a factor of 2 or 3 from one
night to the next and up to 1.5 on the same night\footnote{We note
that a spectrum of the nuclear region in Pictor A which we loaned for
the purpose of measuring line widths has been published with this
faulty calibration and should be disregarded in any absolute
comparisons - see \cite{Su95}.}. Thus it is not
possible to obtain meaningful absolute values for the emission line
strengths nor is it possible to obtain meaningful flux ratios for
emission lines, such as [OI]$\lambda$\,6300 and [OIII]]$\lambda$\,5007,
which were taken with different grating settings on different nights.
We have, however, determined the {\it relative} corrections to the
ratios, using the published absolute flux values
for the stars observed.  Even in the most extreme case of
[O\,I]/H$\alpha$ these are less than $\pm$\,0.04 in the log. 
A more detailed discussion of the spectral measurement and
calibration techniques used for the entire set of CTIO spectra can be
found in (\cite{Gr99}).

\subsubsection{Velocity Analysis}

The velocity--position data from different emission lines in the same
PA were combined and fit with a spline curve.  This spline fitting
technique gives 1) a mean central redshift of 10548 $\pm$21 km$s^{-1}$
and 2) rms deviations from the fitted curves at any one position angle
which range between 22 and 25 km$s^{-1}$. There are some cases in the
literature where the nuclear velocities for AGNs obtained from [OIII]
($\lambda \lambda$4958/5007) are systematically blueshifted with
respect to those obtained from measurement of other optical emission
lines (\cite{Mk94}, \cite{wil85}). However, the consistency between the
rms errors for the nuclear velocity from different position angles, and
the errors obtained from the spline fits at any one position angle,
suggest that this is not a significant factor for Pictor A.
Figure 4 (a-h) shows the original velocity measurements for each PA plotted on
the right hand side and the resulting composite curves, zeroed to the
nuclear velocities, on the left.

\placefigure{fig4}

Although the velocity curves in Figure\,4 appear to show a
rotational trend in the ionized gas (receding most rapidly toward the
NE and approaching in the SW), the sharp excursions from this trend are
as notable as the trend itself. In particular, the velocities in PA
90$^{o}$, 102$^{o}$, and 115$^{o}$ have a wave--like appearance with an
amplitude of 50 to 100 km s$^{-1}$ (also seen in PA 290$^{o}$,
\cite{Ro87}.) To help clarify these peculiar velocities, we have
constructed a ``velocity map'' of the ionized gas in the inner regions
of Pictor A.   Figure 5a shows the spectrograph slit positions
projected onto the optical continuum image of the galaxy.  Since the
slit widths (1.2$\arcsec$) and the separation between pixel columns
(0.73$\arcsec$) were less than the effective seeing profile
($\sim$\,1.4$\arcsec$ FWHM), adjacent positions in a map of the velocity
plotted on the sky for each slit position contain correlated velocity
information.  The spline fits to the emission line velocities for
different position angles were used to generate such a two--dimensional
``velocity map'' of the ionized gas. In areas where the slit positions
were within the region of overlapping seeing, this redundant
information was used to interpolate points in the ``map.''

\placefigure{fig5}

The result is shown in Figure\,5b. There, negative velocities
are shown as dashed contours (ranging from -70 km$s^{-1}$ to -10
km$s^{-1}$ in steps of 15  km$s^{-1}$) and positive velocities are
shown as solid contours (ranging from +10 km$s^{-1}$ to +100
km$s^{-1}$, again in steps of 15 km$s^{-1}$). The technique used does
smooth the velocities in the map, but by what appears to be no more
than $\sim$ 15 km$s^{-1}$. The gaps in the velocity information for PA
15$^{o}$--30$^{o}$, 60$^{o}$--90$^{o}$ and 128$^{o}$--180$^{o}$ are
apparent. However, the general trend seen in the individual line plots
is clear. 1) There appears to be a null, or projected rotation axis at
$\sim$120$^{o}$ -- 140.$^{o}$  2) There is a redshifted velocity plume
on the {\it blueshifted side} of this null region, roughly
1.5--2.5$\arcsec$ to the W of the nucleus (in a cone between PA $\sim$
270$^{o}$ and $\sim$ 290$^{o}$, marked as +45 km$s^{-1}$). This is the
feature which shows up so strongly in the individual velocity curves
for PA 90$^{o}$, 102$^{o}$ and 115$^{o}$ (Figure\,4e--g) and
(from those individual plots, where smoothing is less important) has an
amplitude in excess of 60 km $s^{-1}$ in PA 102$^{o}$.

\subsubsection{Analysis of Emission Line Intensities}

\begin{enumerate}

\item {\it The Gas Along the Velocity Plume:\,\,}
To investigate the physical conditions in the gas which gives rise to
the redshifted velocity feature on the W side of the nucleus, we
measured the relative intensities of the {\it narrow part} of the
emission lines from the spectra in PA 90$^{o}$, 112.5$^{o}$, 115$^{o}$
and 128$^{o}$. The region of interest lies between 1$''$ and 2.5$''$
from the nucleus in each of these position angles and covers 3 to 4
columns in the ccd images. Figure 6 shows a plot of the relative flux
from the nucleus for the red spectrum in PA 102$^{o}$ and the region
around the H$\alpha$ line from the nucleus and 3 ccd columns towards
the W. As noted above, the space between columns is less than the FWHM
of the seeing profile, and these spectra are not independent. This is
particularly evident for the H$\alpha$ line (Figure\,6), where
the very broad component (presumably from the nucleus) can still be
seen at 1.4$''$ to 2$''$ from the nucleus. Nevertheless, by isolating
the narrow line component, we should be isolating the emission from the
low--density gas.

\placefigure{fig6}

The only position angles covering this redshifted velocity peak which
were observed in {\it both} the red and the blue spectral region are
those in PA 102$^{o}$ and 128$^{o}$.  The line ratios from these
spectra (uncorrected for reddening) for the emission lines  [OIII]
(5007)/(H$\beta$), ([SII] 6716+6731)/(H$\alpha$), ([NII]
6583)/(H$\alpha$), and ([OI] 6300)/(H$\alpha$) are shown in
table~\ref{tbl-3}. We have observed stellar flux calibration standards
for all of these spectra and have used them to estimate the {\it
relative} corrections to the ratios in table~\ref{tbl-3}.  Even in the
most extreme cases ([O\,I] : H$\alpha$) these are less than $\pm$\,0.04
in the log (much less than the errors from ``noise'' for the emission
at distances greater than 0.2$''$ from the nucleus - which are
$\pm$\,0.15 in the log). We reiterate here the important observational
point made by \cite{VO87}: {\it Absolute} flux measurements are very
difficult to acomplish, while diagnostic {\it line ratios} for adjacent
emission lines taken simultaniously can be very accurate. The
correction for differential atmospheric refraction between
[O\,III]--$\lambda$5007 and [O\,II]--$\lambda$3727 is well known to be
in excess of 0.3$''$ even for ``reasonable'' zenith angles and even for
one spectrum with the exposure times reported here. When transferred
{\it between} objects (standard star to galaxy) this can easily amount
to a 30\,\% error (under ideal conditions with an ideal spectrograph).

\item {\it The N Optical--Radio ``Pseudo--Ring'':\,\,} The most
mysterious feature seen in the optical--radio images is the large arc
or loop to the NW (at $\sim$ 4$''$ in PA $\sim$ 255$^{o}$), where the a
bubble in the ATCA, 3\,cm image crosses the H$\alpha$ pseudo--ring
(Figure\,1b inset.)  We have no spectra which cover this
position angle directly. The closest coverage is along the spectrum in
PA 0 $^{o}$.  For completeness, we have also measured line ratios for
this spectrum. These are included in table~\ref{tbl-3}.

\placetable{tbl-3}

\item {\it Diagnostic Diagrams:\,\,}
The line ratios are plotted in Figure\,7 (a-c) as diagnostic
``BPT'' diagrams (\cite{BPT81}, \cite{VO87}). The solid curve in each
case represents the division between HII region like objects and narrow
emission--line galaxies such as Seyfert 2, NLRG, and ``liners'' (as
indicated by the empirical delineations in \cite{VO87}.) The ([NII]
6583)/(H$\alpha$) measurements appear to be inconsistent with the [OI]
and [SII] measurements. The latter clearly lie to the right of the
curve delineating the HII region-galaxy separation while the [NII]
measurements lie to the left of the corresponding curve in the  [OIII]
(5007)/(H$\beta$)---([NII] 6583)/(H$\alpha$) plane. The reality of
this effect can be seen in the line plots shown in Figure\,6,
where the nuclear spectrum in the region of H$\alpha$--[NII] from PA
102$^{o}$ is shown plotted to the same scale as the corresponding
spectra at 0.73, 1.46, and 2.25$''$ W of the nucleus. (The relative
scaling amounts to roughly a factor 8 between the nuclear spectrum and
that at 2.25$''$ W.) The redshift  (seen as the redshifted velocity
plume in Figure\,5b) clearly shows up in the narrow line region
of both the H$\alpha$ and the [NII] lines. The single large circle and
square plotted in each diagram refer to values of line ratios from a
model by \cite{Fe97} and will be discussed in section 4.

\placefigure{fig7}

\item {\it [S\,II] Line Ratios:\,\,} Finally, we determined the [SII]
(6717/6731) ratios for the regions in question (table~\ref{tbl-4}), as
a check on our assumption that the measured lines come from the
low--density region outside the nucleus.  These do show a slight
tendency to decrease with distance from the nucleus (corresponding to a
density change in $N_{e}$ from roughly 10$^{3}$ to $10^{2}$ (with T =
10,000 $^{o}$ K) (\cite{Os89}). We note, however, that although the
{\it trends} seen in the densities inferred from these [S\,II] emission
lines are probably real, the {\it densities} probably cannot be
derived directly from the measured line ratios (cf. \cite{Fe97}.)

\end{enumerate}

\placetable{tbl-4}

\section{HST Observations } 

The primary goal of our HST observations was to  measure the size of
the narrow-emission-line region (NLR) surounding the nucleus, and to
search for morphological features (such as a compact, double nucleus,
or an optical jet related to the radio emission) which might provide
clues about the mechanisms feeding the nuclear energy source.  Current
physical models predicted that the nucleus of this object should be a
point source and contain roughly 90\% of the total flux observed from
the ground within 1\arcsec \/.  Our CTIO spectral measurements (above)
yield a lower limit to the total flux in the [O\,III]
$\lambda$5007/4958 lines of $10^{41}h^{-2}$ erg s$^{-1}$.  Using this
and standard parameters for the emitting gas clouds, we estimated the
scale of the transition region between narrow--line and broad--line
emission to be 75--150\,pc (or  $\sim$0.15 $''$$h$  to $\sim$0.30
$''$$h$, roughly 3.5 to 7 pixels at 43 mas pixel,$^{-1}$ the scale of
the original HST Planetary Camera.

\subsection{The Choice of Camera and the Undersampling Problem}

We chose to observe Pictor A with the Planetary Camera (PC)
(\cite{wes82}) because the ability of the CCD detector to accept high
photon count rates would minimize observing time and maximize signal to
noise in the fainter, diffuse regions adjacent to the nucleus.  In
addition, the filters available on the original WF/PC included one
which would isolate the redshifted [OIII] emission (F517N) and this
presented an excellent opportunity to image one of the more important
NLR components for this powerful, nearby FRII radio galaxy.  When it
became clear that the original HST camera/mirror suffered from
spherical aberration we decided to go ahead with a subset of the
proposed observations on Pictor A because we believed (correctly, it
turned out) that any subsequent repair would eliminate the F517N filter
and thus rule out the possibility of imaging the [OIII] region.  With
the aberrated images found in Cycle 1 and  Cycle 2 data, it was
necessary to increase the observing time by a factor of 7 over
unaberrated exposures to compensate for the increased photon noise from
the scattered light in the diffuse regions surrounding the nucleus.

\subsection{The HST Data}
During Cycle 1, Pictor A was observed through the F517N and F547M
filters on Sept.\ 12, 14, 15, and Oct.\ 8, 1991 (table~\ref{tbl-3}).
The F517N filter was centered on the  $\lambda\lambda$ 4958 and 5007
emission lines from [OIII] while the F547M filter includes these
emission lines at its blue end, with roughly the same transmission, but
also transmits approximately five times more continuum emission than
the F517N filter (\cite{WF89}).  Our objective was to separate out the
NLR from the continuum emission by subtracting the different filter
images.

During Cycle 2, Pictor A was observed through the F648M and  F718M
(\cite{WF89}) filters on Sept.\ 20, 1992 (table~\ref{tbl-3}).  The
F718M filter covers the emission lines from H$\alpha$, $\lambda$ 6563,
[NII] $\lambda\lambda$ 6527, 6548, 6583\/ and [SII] $\lambda\lambda$
6717--6731, while the F648M filter includes both the [OI] emission
lines at $\lambda\lambda$ 6300, 6363 which appear in the ground-based
nuclear spectrum at a factor of $\sim$ 20 less intensity than the
H$\alpha$ emission line (cf. the inset in Figure\,6). The F648M
filter  also includes the H$\alpha$ and [NII] lines, but at a reduced
transmission of roughly a factor of 2.5.  Our objective here was to
separate out the H$\alpha$ + [NII] emission line regions from the
continuum emission, although the filter combination was far less
favorable for this than the one isolating [OIII] in Cycle 1.

The exposure times for these observations were calculated from the
published transmission--detector curves for the HST PC (\cite{WF89})
and our nuclear spectra for Pictor A (described above). The
difficulties for this type of calculation are two-fold: 1) The filter
transmissions change with time and temperature, and transmission values
for wavelengths near the filter edge (such as the [NII], H$\alpha$
group in the F648M filter) may vary significantly, and 2) The nuclear
emission line spectrum of Pictor A is known to vary
(\cite{Fp85,Su95,Ha96}) and the change in flux can (and evidently did -
see section 3.3.3) lead to saturated images.

Both the Cycle 1 and the Cycle 2 observations had the obvious drawback
that the true, observed PSF for this object would have an ``effective
filter'' which was a function of the relative emission-line strengths
at each point and their redshifted wavelength. However, this is a
secondary effect compared with the aberration problem.

\subsection{HST Data Analysis}

All of the observations were normalized and bias corrected using the
most proximate calibration frames in the data base. They were cleaned
of cosmic ray ``hits'' using standard STSDAS WFPC analysis routines.

Since the HST primary mirror has a diffraction-limited resolution of
57.7 mas at 550 nm and 37.8 mas at 720 nm, the PC data, with a grid
size of 0.042 mas pixel$^{-1}$ is not properly sampled. Thus, to
construct a properly sampled image for our Cycle 1 data
 we chose to ``dither'', i.e., observe the object at four slightly
shifted positions (offset in $\Delta x$ and $\Delta y$ by different
pointing combinations of 0.022 mas as shown in Figure\,8a.)

\placefigure{fig8}

In fact, because of scheduling constraints, this tidy scheme was not
used and the observations for Cycle 1 were offset by random pixel
shifts, at different spacecraft roll angles. The offsets were measured
using two methods; 1) crosscorrelation of individual images taken
through the same filter (using the 14/09/91 data as a reference) and 2)
using the ``imcnt'' routine in IRAF to measure the pixel coordinates of
the bright, nuclear region. These methods gave results which agreed to
within less than 0.05 pixel. The measured pixel shifts are given in
table~\ref{tbl-3} (and shown graphically in Figure\,8b). In
addition, there was a systematic shift between the images taken through
the F517N filter and the F547M filter. This amounted to --0.90 $\pm$0.06
in x and +0.10 $\pm$0.05 in y (in good agreement with independently
measured values of --0.69 and --0.05 by \cite{Ev91}. The F517N images
were shifted by this additional amount after deconvolution.

\subsubsection{Analysis of Composite Images}

Many different schemes have been developed for ``dithering'' image data
before ''cleaning'' restoring it with a more visually interpretable
PSF. Several of these and their applicability to HST Cycle 1 and Cycle
2 data are described in \cite{Ha94}. Because our Cycle 1 data is
unique, we outline the scheme we used here:

A rotated image is the same as a non-rotated image convolved with a PSF
of opposite rotation.  That is, if $I_{i}$ is an image at roll angle
$i$ and $e_{i}$ is its exposure time, then:

\begin{equation}
(I_{1}\ast P)_{\theta} + (I_{2}\ast P)_{\theta} + (I_{3}\ast
P)_{\theta} + (I_{4}\ast P)_{\theta} = I_{\theta} \ast ( \sum
P_{-i} \cdot e_i)
\end{equation}

where $I_{\theta}$ and P are the image and PSF at ${\theta}^{o}$, the
subscript $\theta$ outside the parentheses means the image has been
rotated back to ${\theta}^{o}$, and $P_{-i}$ is the PSF at ${\theta}^{o}$ \/
rotated through the negative roll angle, $-i$.
Thus, 

\begin{equation}
P_T \equiv (\sum P_{-i} \cdot e_i)
\end{equation}

becomes a new, composite PSF for the composite image which is the sum
of the individual images shifted to a common center and
rotated back to a common position angle, $\theta$.

Note that since the rotation operation can be expressed in terms of
two orthogonal functions (sine and cosine), as long as two conditions
hold: a) the undersampling is less than a factor of 2, and b) for each frame,
if $\Delta p$ is the pixel size, and both the offsets from the common 
center, $\Delta x$ and $\Delta y$, include a pixel fraction, $f_{p}$, such that

\begin{equation}
0.25\Delta p  \leq f_{p} \leq 0.5\Delta p \; ,
\end{equation}

then all of the pixels in the registered images will have sufficient offset
spacings to sample a different part of the image.
Both these conditions hold for the Cycle 1 data discussed here. 

One advantage of this serendipitously employed technique is that the
composite PSF, $P_T$, is much more radially symmetric, with the ``tendrils''
de-emphasized (and thus, their propensity to induce artificial, radially
extensive features minimized).

\section{The Restored HST Images}

\subsubsection{Cycle 1 Results}

The technique described above was applied to our Cycle 1 data which was
resampled on a double grid, shifted to a common center, rotated using a
sinc interpolation, and then co-added. One set of PSF's used for
deconvolution was generated in the same way starting with a set of {\it
observed\/} PC6 PSFs at 502nm (supplied by Ed Groth and Jim Westfall,
and regridded to reflect the slight change in the PSF between 502nm and
517nm). These were slightly shifted from the PC6 position of the galaxy
nucleus but, again, this (along with the slight mis-match in effective
wavelength of the PSFs) is a secondary effect. A second set of PSFs was
generated using the ``Tiny Tim'' optical modeling software
(\cite{bur91}; \cite{Kr94}). The relevant central parts of these two
PSFs are shown in Figure\,9a and b. Their ratios (observed /
calculated) are shown in Figure\,9c. The principal difference
between the two PSFs is the fine detail in the area around the central
pixel which is found in the ``Tim'' PSF and not in the ``observed''
one. This amounts to a 10\/\% amplitude difference and appears to
influence the final deconvolution result (see section 3.3.2.)

\placefigure{fig9}

The composite images were restored using both of these different sets
of point-spread functions and  a Maximum Entropy algorithm, with a
front end (MEM) developed by Weir (1991a; 1991b).  Since the strong
[OIII] emission lines from the nuclear region of Pictor A are included
in both the F517N and F547M band passes, the PSF structure will be
determined primarily by these narrow emission lines, which are better
approximated by the F517N PSF. Thus the approximate F517N PSFs were
used for deconvolving both the observations through the F547M filter
and those through the F517N filter.

The discussion above makes it clear that we do not have a perfect,
mathematically correct, set of data for deconvolution analysis.
Nevertheless, we have taken an experimental approach, doing scores of
deconvolution calculations, using different combinations of what seem
to be ``reasonable'' approximations to the PSFs, image subtraction
procedures, and deconvolution criteria. All of these different
calculations produce images which share the same general features. 

The two most extreme results for the two filter sets (F517N and F547M)
are presented here.  These two extreme cases differ in deconvolution
procedures in two important ways: 1) one set (shown in
Figure\,10 a and c and 11 a, b, and c) uses the ``observed''
PSFs while the other (Figure\,10 b and c and 11 d) uses the
``Tim'' PSFs and 2) the registration and construction of the raw images
and PSFs for the images in Figure\,10 a and c and 11 a, b, and
c  was done in such a way that North was towards the left (decreasing
x) side of the frame while the images shown figures 10 b and d and 11 d
were oriented with North pointing at -21.72$^{o}$ (i.e.  clockwise)
from the increasing y axis. (All images have been rotated to the normal
astronomical coordinate system in figures 10 and 11.) These different
orientations were chosen to examine the influence of the coordinate
grid orientation on the final deconvolution result. It proved to be as
important as the difference in PSF images.

\placefigure{fig10}

The results for the F547M images are shown in Figure\,10a-d.  Both
reconstructions give approximately the same normalized peak flux for
the nucleus (11,800 using the ``observed'' PSF with results shown in
Figure\,10 a and 13,500 using the Tiny Tim PSF, with results
shown in Figure\,10 b). The images in Figure\,10 c and
d have been corrected for the [OIII] emission by subtracting off 54\%
of the flux in the F517N images (the 54\% contribution was calculated
using the filter/instrument response curves in the \cite{WF89}.
These values clearly are not stable, but are probably no more in error
than any of the other assumptions and approximations used in this
analysis. The  Tiny Tim PSF yields images which are consistently more
``point like'' than those deconvolved with the observed PSF. This seems
to be related to the difference in PSF structure shown in
Figure\,9 c. We conjecture that the fine detail in the
off--center structure of the Tim PSF leads the deconvolution
algorithms to treat the measured photons in extended, smooth
structures as ``noise'', forcing them into any adjacent point sources.
Both reconstructions show what appears to be real structure outside the
nucleus.  

In particular, there is evidence for a small continuum feature in the
direction of the large--scale radio structure (shown as arrows in
Figure\,10 c and d, corresponding to the arrows shown in
Figure\,2 inset) with a peak flux which is $\sim$4\/\% of that
in the unresolved nucleus in Figure\,10 a and $\sim$3\/\% in 10
c with corresponding values of $\sim$3\/\% and $\sim$2\/\% for the
images in figures 10 b and d, respectively. This feature consistently
showed up in all of our F547M reconstructions with peak intensity
values ranging from 22:1 to 55:1 relative to the nucleus. This feature
is certainly continuum emission, not emission from ionized gas. (It
does not show up in the [OIII] image, Figure\,11.) It could be
a faint star or scattered light from the nucleus but the probability of
such an accidental alignment with both the outer radio structure and
the VLBI jet (see section 3.5, below) is quite low. The most likely
possibility is that it is associated with the radio structure in Pictor
A.

\placefigure{fig11}

The deconvolutions for the F517M images using the ``observed'' PSF are
shown in Figure\,11 a and b.  Figure 11 a shows a gray--scale
plot of the deconvolved F517N image (uncorrected for continuum).
Isophotes for the deconvolved F547M image (obtained with the same PSF)
are superimposed.  Figure 11 b shows the deconvolved F517N image
plotted alone. Note the broad plateau to the N of the nuclear peak.
Figure 11 c is the deconvolved F517N image (observed PSF) corrected for
continuum by subtracting off 30\/\% of the deconvolved F547M image
(again, the relative [OIII] -- continuum contributions to the two
filters are based on the data from the \cite{WF89}). Finally,
Figure\,11 d shows the result of deconvolving the {\it
difference} between the two observed images with the ``Tiny Tim'' PSF
(i.e. subtracting 30\% of the F547M image from the F517N image with the
grid oriented as it was for the images in figures 11 b and d and then
doing the deconvolution). The most notable features in these images
are: 1) In both (very different) deconvolution results the [OIII]
emission is extended by $\sim$ 150 mas to the NW, {\it outside} of the
path of the ``jet'' or ``bridge'' (again shown by the arrows.) 2) In
both images (Figure\,11 c and d) the [OIII] emission peaks at a
point located $\sim$ 60 mas to the {\it NE} of the nuclear position.

\subsubsection{Cycle 2 Results}

We obtained two further sets of images with the HST PC in Cycle 2,
using medium--width filters centered at 648\,nm (F648M) and 718\,nm
(F718M).  As noted above in section 3.2), the F718M filter covers the
emission lines from H$\alpha$, $\lambda$ 6563, [NII] $\lambda\lambda$
6527, 6548, 6583 and [SII] $\lambda\lambda$ 6717, 6731, while the F648M
filter includes both the [OI] emission lines at $\lambda\lambda$ 6300,
6363.

The Cycle 2 exposures, were scheduled and taken before the Cycle 1 data
had been analyzed, and were designed to obtain maximum signal-to-noise,
without ``dithering'' in the allotted spacecraft time.  The images were
all recorded on the same place on the PC 6 chip (to within 0.05 pixel)
and with the same roll angle.  Thus  the images could  not be shifted
and treated as a fully sampled image set, making it impossible to
recover the full resolution of the telescope as was done with the Cycle
1 images.  In addition, the F718M images were saturated in the nucleus,
suggesting that this filter contained more of the broad nuclear
H$\alpha$ emission line than we had calculated (probably because the
flux in the  H$\alpha$ line had increased since our CTIO observations,
\cite{Fp85,Su95,Ha96}.)

Nevertheless, these Cycle 2 images were also double binned and
co-added.  The image in Figure\,12a is an attempt to smooth the
F648M image to a resolution of $\sim$ 1$''$ for comparison with our
ground--based H$\alpha$ image. However, there are not sufficient
detected photons in the F648M frames to provide a reliable,
low--resolution image. The Weir ``MEM'' restorations were done for the
two filter data sets using ``Tiny Tim'', composite PSFs at the correct
wavelength. The results for the F648M filter are shown in
Figure\,12b. (The results for the F718M images are not worth
publishing.) It is clear that this reconstruction displays no extended
structure beyond that expected from the distortion of the aberrated PSF
itself, and no attempt was made to do the type of filter subtraction
which was successful for the Cy 1 data.  The most logical explanation
for this ``failure'' is simply the fact that the data for Cycle 2 are
undersampled (and saturated for the F718M filter data).

\placefigure{fig12}

\subsection{Comparison of the HST Images with a VLBI Core Jet} 

After our initial reductions of our Cycle 1 HST data (in June of 1992),
when we had identified the continuum knot or ``jet'' seen in
Figure\,10a  (and before \cite{Pr97} had firmly concluded that
there was a ``bridge'' linking the NW hotspot to the core in their
20\,cm data), we learned that the 2.291 GHz data from the SHEVE  VLBI
array (\cite{Ti99}), taken roughly half a year before our HST Cycle 1
data, showed evidence for an extended structure in the ``E-W''
direction.  Subsequently, two additional sets of SHEVE data, at 8.4
GHz, were obtained (in February and July, 1993, \cite{Ti99}.)

Two of the SHEVE images (2.291 and 8.418 GHz) are over
plotted on one of our HST F547M restorations Figure\,13.m (The
2.291 GHz data defines the outer contours of the inset image and the
8.418 data is shown as gray--scale.) The HST ``continuum'' image
plotted in Figure\,13 has been corrected for [OIII] emission (as
described in section 3.4.1). It differs from the restorations shown in
figures 10 and 11 in the way the gridding was done before
deconvolution. The HST image in Figure\,13 was restored with a
``Tim'' PSF (see section 3.3.2) but the prerestoration grid was
oriented in the normal (E to left N to top) astronomical system. The
``boxy'' nature of the central isophotes which seem to be associated
with this grid orientation can be seen in Figure\,13.

Although the SHEVE images have too low a dynamic range to identify any
direct correlation between the HST continuum image knot and the radio
structures (\cite{Ti99}), the alignment is certainly suggestive (as is
the overall relationship between the VLBI jet, the HST continuum, and
the HST [OIII] image morphology -- see section 4 below.)

\section{Discussion}

This compendium of data marks the first time that both optical and
radio measurements have been available at such high spatial resolution
for a well--defined FRII radio galaxy.  Although there is a wealth of
details in the data, our objective in this paper is to focus on the
features in the different data sets which  enhance our understanding of
the relationship between the beamed radio plasma and the ambient
interstellar medium in such galaxies. The following points appear
significant:

\subsection{Relationship Between the Small-scale and Large-scale Morphology}

Within the resolution of our HST images, the spatial relationship
between the radio and optical features on different scales appears to
be homologous.

The most obvious morphological similarities are those involving
alignment.  The VLBI jet, the HST continuum feature,  the end of the
H$\alpha$ ``bridge'' in PA 283$^{o}$, the 20\,cm VLA ``bridge'', and
the direction between the nucleus and the outer hot--spots all lie
within the range PA 280$^{o}$ -- 288 $^{o}$.

A more subtle morphological similarity seems to occur in the offsets
between the continuum and emission features. On the pc--mas scale, the
optical {\it continuum} feature and VLBI jet are nearly aligned but the
[OIII] line emitting region is extended (by roughly 200 mas or 110
pc$h^{-1}$) to the NW (Figure\,11 b and d.). On the kpc scale,
the VLA bridge and the direction to the NW hot--spot are aligned but
the H$\alpha$ bridge,  which extends to the NW,  lies slightly to the N
of a line linking the nucleus with the jet--hot--spot direction. In
both cases, on very different scales, the line emitting gas appears to
the N of the channel defined by the radio beam.

There is also an asymmetry in the central brightness of this object
which shows up on both the 100 pc and the kpc scales. Referring back to
Figure\,5a, the steepness of the continuum intensity gradient
on the S side of the galaxy also shows up in the H$\alpha$ image
(Figure\,5b), where the nucleus of the galaxy (at 0,0 in the
plot) lies to the S of the brightest H$\alpha$ emission. Turning to the
plots in figures\,11c and d, we see that there the brightest
[OIII] emission also lies to the N (and slightly E) of the nucleus.
This asymmetry {\it may} be caused by dust in the nucleus (cf.
\cite{Kn90} for a discussion of the IRAS data) but, again, it may be
associated with the asymmetry in the velocity field and the 3\,cm ATCA
feature seen to the N of the nucleus (below).

Finally, and most puzzling, the Kpc scale ``pseudo--ring'' and 3\,cm
feature seen to the N of the optical nucleus seem to have a
small--scale counterpart in the peri--nuclear [OIII] emission found
with the HST. Although the resolution of the latter image is marginal,
it appears as a ring--like feature with an enhanced brightness to the N
in figure 11 c and as an extended envelope with a point--like
enhancement to the N in figure 11 d. In this case, the emitting gas
seems to take on a ring--like structure at very different scales.

The overall impression which one receives from these morphological
homologies is that the radio jet or beam is pushing the ambient
interstellar gas to the N in its passage out from the nucleus.

\subsection{Orientation of the Velocity Structure in the EELG}

Two characteristics of the velocities seen in figures 4 a--h and 5 b
imply a correlation between the kinematic features of the EELG and the
beamed radio plasma:

\begin{enumerate}

\item The initial impetus for our CTIO spectroscopy program was to
obtain sufficient information about the velocity field in the EELG to
separate out large--scale rotational motions (if any) from more chaotic 
flows in the gas. Although we do not have complete coverage of the
galaxy, there does appear to be a ``null'' in the overall large--scale
velocity field at $\sim$\,PA 285$^{o}$--PA 295$^{o}$. This is
consistent with the picture that well--defined FRII radio galaxies have
radio axes which are correlated with the projected rotation axes of
their gas (\cite{Si79}, \cite{He85}.)

\item Of more direct relevance to the question of interaction between
the radio plasma and the ambient interstellar gas is the red--shifted
velocity ``plume'' seen in the direction of the jet--bridge--radio lobe
alignment (Figure\,5b.) Although the relative projected
velocities are small (the maximum deviation from ``circular velocity''
in Figure\,4f seems to be $\sim$ 60 km$s^{-1}$), this lends
strong support to the morphological evidence that the radio plasma has
``clobbered'' the interstellar gas. There is, however, the alternative
possibility that the  counter velocities seen on the W side of the
nucleus are not the result of disturbance by the radio plasma but are,
instead, a gravitationally induced flow in response to a
non-axisymmetric potential. This latter type of velocity disturbance
would most likely have a signature which showed up as a counter flow in
$\sim$ PA 30$^{o}$--90$^{o}$ where our spectral coverage is poor. Thus
it cannot be completely ruled out.

\end{enumerate}

\subsection{Emission Line Ratios}

The most outstanding feature of the emission line ratios in
table~\ref{tbl-3} and Figure\,7 is the very low value of
[NII]\,:\,H$\alpha$.  Although the [SII]\,:\,H$\alpha$ and
[OI]\,:\,H$\alpha$ plots (Figure\,7a and b) appear to place
these regions in the upper envelope of the ``AGN region'' (shown in
\cite{VO87}), the [NII]\,:\,H$\alpha$ ratios (Figure\,7c)lie in
the area of the diagram occupied by HII  regions. The AGN spectra which
appear to bear the most similarity to the spectrum of Pictor A in
Figure\,6 are those of Seyfert 1.5 and 1.8 galaxies. However,
the latter objects do not have the relatively high ratios of
[OI]\,:\,H$\alpha$ and low ratios of [NII]\,:\,H$\alpha$ seen here
(\cite{Co83}).

There are reports in the literature of EELG with low
[NII]\,:\,H$\alpha$. Most commonly, these seem to  gas clouds which are
clearly associated with or entrained in a radio jet. The data for these
objects has been summarized by \cite{Vi92}.  More recently, the EELG
associated with the hot--spot and  SE radio lobe in PKS 1932-464
studied by \cite{Vi98} and for PKS2250-41 by \cite{Cl97} have also been
observed. In all of these cases, however, although the reported values
for  [NII]\,:\,H$\alpha$ and [OI]\,:\,H$\alpha$ are lower than those
usually found for active galaxy nuclei, none of the examples are as
extreme as the ones we have found for Pictor A (including its nuclear
values).

As noted earlier, there are competing models for calculating the
expected emission line ratios from shock heating and/or UV continuum
excitation and ionization for the EELR in galaxies as well as the
nuclear spectrum of AGNs. \cite{Bab89} give a good summary of the field
up to that date. More recently, additional, more complex calculations
have been carried out (\cite{Vi92,Do95,SB96,Fe97,Al98} and references
therein.)

Although all of these models do manage to fit the observed line ratios
found in HII regions and AGN spectra, none of them, with the exception
of models recently published by \cite{Fe97}, reproduce the extreme
values of the [OI] and [NII] ratios reported here. The predicted line
ratios from \cite{Fe97} which do seem to fit our present values are
plotted as large open circles and squares for the diagnostic planes in
Figure\,7. These correspond to models in which the covering
factor for clouds drops off much more slowly with radial distance from
the ionizing source than is the case for models which fit the observed
spectra of most AGNs (``Dusty'' models where the cloud covering
fraction falls off roughly as r$^{-1.25}$ and n$^{-1.0}$ and ``solar
abundance'' models where it falls off as r$^{-0.75}$ and n$^{-1.0}$,
\cite{Fe97}).

However, it is difficult to reconcile the reasonably good fits to some
of the \cite{Fe97} spectral models with the geometry of the EELG which
we observe in Pictor A. Although the gas in the regions 2$''$--3$''$ (1
to 1.6 kpc$h^{-1}$) may well receive ionizing radiation from the
nucleus, the fact that strong [OI] emission is found only in the
non--nuclear region associated with the velocity ``plume'' and the
direction of the H$\alpha$ and VLA bridge suggests that there must also
be some energy input to the gas from either shocks or cosmic rays
associated with the radio plasma.

\section{Conclusions}

By combining a series of high angular resolution observations at
multiple frequencies, we have been able to establish a fairly
convincing case for the direct physical disruption of the interstellar
medium in the inner regions of a classical FRII radio galaxy. In
particular, we find morphological features which suggest that the
structural disruption extends over scales ranging from a few hundred
parsecs to several kpc.  One of the spectral signatures of this
disrupted gas appears to be anomalously weak line strengths for [NII]
compared with H$\alpha$. Although the low emissivity from the [NII]
line may reflect intrinsically low N abundances in this particular
radio galaxy, it seems likely that it may be related to the excitation
mechanism for these gas clouds.

All observational papers end by concluding that more and better data
are needed to answer the questions raised by the data reported. In this
paper we wish to vary that theme somewhat. What is needed, if we are to
achieve a better understanding of the physics of radio jets and their
ambient medium is not {\it more } data but {\it different} data. In particular, we emphasize the importance of obtaining high resolution measurements (both spatially and spectroscopically) of the entire velocity field in the EELG of nearby radio galaxies.
 Only by obtaining very high angular resolution measurements for the very closest objects will we have sufficient spatial resolution to permit sensible comparison with spectral models.

\acknowledgments

This work is based on observations with the NASA/ESA Hubble Space
Telescope obtained at the Space Telescope Science Institute, which is
operated by Association of Universities for Research in Astronomy,
Incorporated, under NASA contract NAS5-26555 and with financial support
from HST grant GO-245601-87A.  We thank both CTIO (operated by AURA
under contract to the US NSF) and the ATNF (funded by the Commonwealth
of Australia for operation as a National Facility managed by CSIRO) for
observing time and support. Part of this work was done with financial
support from NSF (AST-8914567) partial publication costs came from a
grant (to SMS) from NASA, administered by the American Astronomical
Society).  The graphics in this paper were generated with the software
package $WIP$\footnote{$WIP$ is copyright by the
Berkeley-Illinois-Maryland Association (BIMA) Project \cite{Mo95}},
using the $PGPLOT$\footnote{software copyrighted by California
Institute of Technology} graphics library.  This research has made use
of the NASA/IPAC extragalactic database (ned) which is operated by the
Jet Propulsion Laboratory, Caltech, under contract with the National
Aeronautics and Space Administration. We thank  Ed Growth and John
Mackenty for help obtaining observed PC PSFs, John Kriss for producing
the "Tiny Tim" software, and Ian Evans for help with N.  Weir's memsys
routines.

\begin{deluxetable}{ccccc}
\hspace*{-12.0in}

\footnotesize
\tablecaption{CTIO - 1.5m - Optical Images \label{tbl-1}}
\tablewidth{0pt}
\tablehead{
\colhead{ DATE }  & \colhead{AIRMASS }  & \colhead{FILTER} &
\colhead{EXP TIME} & \colhead{SEEING (FWHM)} 
} 

\startdata
27/28 Feb 1987  &   1.09   &    6826/78 line & 2000 s & 1.7$''$\nl
27/28 Feb 1987  &   1.14   &    6560/110 cont & 2000 s& 1.7$''$\nl
27/28 Feb 1987  &   1.22   &    6826/78 line & 2000 s & 1.7$''$\nl
28 Feb/1 Mar 1987  &   1.46    &    6826/78 line &2000 s &1.5$''$\nl
28 Feb/1 Mar 1987  &   1.69    &    6560/110 cont&2000 s &1.5$''$\nl
\enddata
\end{deluxetable}

\begin{deluxetable}{crc}
\hspace*{-3.0in}

\footnotesize
\tablecaption{CTIO - 4m - Spectra \label{tbl-2}}
\tablewidth{0pt}
\tablehead{
\colhead{ DATE }    & \colhead{SLIT PA }  & \colhead{WAVELENGTH RANGE} 
}

\startdata
 
19/20 Feb 1987  &      90$^{o}$   &      6473 - 7163 \AA \nl  
19/20 Feb 1987  &      12$^{o}$   &      6473 - 7163 \AA \nl                
19/20 Feb 1987  &       0$^{o}$   &	 6473 - 7163 \AA \nl 
20/21 Feb 1987  &     102$^{o}$   &      6473 - 7163 \AA \nl
20/21 Feb 1987  &     115$^{o}$   &      6473 - 7163 \AA \nl                
21/22 Feb 1987  &     128$^{o}$   &      6473 - 7163 \AA \nl
21/22 Feb 1987  &      50$^{o}$   &      6473 - 7163 \AA \nl                 
22/23 Feb 1987  &     102$^{o}$   &      4882 - 5729 \AA \nl
22/23 Feb 1987  &     128$^{o}$   &      4882 - 5729 \AA \nl               
14/15 Jan 1988  &      30$^{o}$   &      6653 - 7338 \AA \nl
15/16 Jan 1988  &      30$^{o}$   &      4895 - 5746 \AA \nl
\enddata
\end{deluxetable}

%\clearpage

\begin{deluxetable}{rccccc}
\hspace*{-5.0in}
\footnotesize
\tablecaption{ Emission Line Ratios  \label{tbl-3}}
\tablewidth{0pt}
\tablehead{
\colhead{PA($^{o}$)}
&\colhead{dist from Nuc ($''$)}
&\colhead{log([O\,I]/H$\alpha$)}
&\colhead{log([N\,II]/H$\alpha$)}
&\colhead{log([S\,II]/H$\alpha$)}
&\colhead{log([O\,III]/H$\alpha$)}
}

\startdata
0& 0.00 \,\,   &  -0.240&   -0.651&  -0.132& \nodata \nl
90& 0.00 \,\,  &  -0.454&   -0.748&  -0.344&  \nodata\nl
102& 0.00 \,\,&  -0.466&   -0.775&  -0.231&   0.461\nl
115& 0.00 \,\, &  -0.518&   -0.726&  -0.242&   \nodata\nl
128& 0.00 \,\, &  -0.563&   -0.706&  -0.334&   0.464\nl
0& 0.73 N&  -0.410&   -0.615&  -0.022&  \nodata\nl
0& 0.73 S&  -0.354&   -0.654&  -0.211&  \nodata\nl
90& 0.73 W&  -0.550&   -0.806&  -0.155&  \nodata\nl
102& 0.73 W &  -0.561&   -0.759&  -0.266&   0.558\nl
115& 0.73 W &  -0.561&   -0.759&  -0.266& \nodata \nl
128& 0.73 W &  -0.521&   -0.743&  -0.259&   0.512\nl
0& 1.45 N&  -0.610&   -0.559&   0.050&  \nodata\nl
0& 1.45 S&  -0.575&   -0.744&  -0.370&   \nodata\nl
90& 1.45 W &  -0.614&   -0.853&  -0.114&  \nodata\nl
102& 1.45 W &  -0.320&   -0.699&  -0.175&   0.516\nl
115& 1.45 W &  -0.504&   -0.819&  -0.465& \nodata\nl
128& 1.45 W &  -0.528&   -0.689&   0.013&   0.476\nl
0&  2.20 N&  -0.579&   -0.551&   0.025&   \nodata\nl
0&  2.20 S&  -0.470&   -0.547&  - 0.225&  \nodata \nl
90&  2.20 W &  -0.640&   -0.727&  -0.048&  \nodata\nl
102&  2.20 W &  -0.182&   -0.588&  -0.114&   0.451\nl
115&  2.20 W &  -0.465&   -0.728&  -0.503&  \nodata\nl
128&  2.20 W &  -0.489&   -0.748&  -0.111&   0.547\nl
0& 3.00 S&  -0.636&   -0.512&  -0.214&   \nodata\nl
\enddata
\end{deluxetable}

%\clearpage

\begin{deluxetable}{cccccc}
\footnotesize
\tablecaption{[SII](6717/6734) \label{tbl-4}}
\tablewidth{0pt}
\tablehead{
\colhead{PA }  
        &\colhead{Nucleus}&\colhead{0.73" W} &\colhead{1.46" W} &\colhead{2.25" W} &\colhead{3.00" W}  }
\startdata

(error)   &    (0.06)&    (0.06)&    (0.06)&     (0.15)&     (0.30)\nl
      90.0&   1.00&   1.25&   1.28&   1.38&   1.11\nl
   102.5&   1.17&   1.24&   1.22&   1.34&   1.65\nl
     115.0&   1.24&   1.21&   1.20&   1.56&   1.39\nl
     128.0&   1.12&   1.23&   0.69&   1.24&    1.19\nl
\enddata
\end{deluxetable}
%\clearpage
 
\begin{deluxetable}{cccccccc}
\footnotesize
\tablecaption{ HST Observations\label{tbl-5}}
\tablewidth{0pt}
\tablehead{
\colhead{HST DATA SET } & \colhead {DATE }   & \colhead{OBS TIME(UT)}   & \colhead{FILTER} & 
\colhead{ $exp_{avg}$}  & \colhead{ ($\Delta$ X)} & \colhead{($\Delta$ Y)} & 
\colhead{P.A. }
} 

\startdata
W00G0201/2T &15/09/91 & 00:10:39/01:31:39& F517N& 598.75s & 2.60& -0.32 &90.499$^{o}$\nl
W00G0203/4T &15/09/91 & 03:04:39/03:17:39 & F547M&304.50s & 2.60& -0.32  &90.499$^{o}$\nl
W00G0101/2T &14/09/91 & 01:37:39/02:58:39 & F517N &  559.00s & 0.00 & 0.00& 91.357$^{o}$ \nl
W00G0103/4T &14/09/91 & 04:31:39/04:44:39 & F547M &  352.25s & 0.00&0.00 & 91.357$^{o}$\nl
W00G0301/2T &12/09/91 &19:01:39/22:22:39 & F517N&  413.50s&1.25& -2.20&  92.515$^{o}$\nl
W00G0303/4T &12/09/91 & 21:55:39/22:08:39 & F547M &  400.00s&1.25& -2.20& 92.515$^{o}$\nl
W00G0401/2T & 08/09/91 & 19:40:39/21:19:39 & F517N & 600.00s&1.70 &1.20  & 68.280$^{o}$\nl
W00G0403/4T &08/09/91 & 22:56:39/23:09:39& F547M &  287.75s & 1.70& 1.20 & 68.280$^{o}$\nl
W11Z0101/2T & 20/09/92 & 05:47:38/07:23:38 & F648M& 1922.50s &  0.00&  0.00&  94.030$^{o}$\nl
W11Z0103/4T & 20/09/92 & 08:34:38/09:06:38 & F718M & 893.50s & 0.00 & 0.00  & 94.030$^{o}$\nl
W11Z0105/6T & 20/09/92 & 10:34:38/12:10:38 & F648M & 1900.00s& 0.00 &  0.00  & 94.030$^{o}$ \nl
W11Z0107/8T & 20/09/92 & 13:32:38/13:55:38 & F718M & 1000.00s0& 0.00 &  0.00  & 94.030$^{o}$ \nl 
\enddata
\end{deluxetable}
\clearpage

\begin{thebibliography}{}
\bibitem[Allen  et al. 1998]{Al98}Allen, M., G., Dopita, M., A.,
Tsvetanov, Z., I., and Southerland, R., S., 1998, \apj, 493, 571
\bibitem[Australia Telescope 1992]{AT92} Australia 
Telescope 1992, J. Electr. Electron. Eng. Aust., Special
Issue, 12, No. 2, editors, R.H. Frater and J.W. Brooks
\bibitem[Baldwin  et al. 1981]{BPT81} Baldwin,J.\/A., 
Phillips, M.\/M., and Terlevich,R., 1981, \pasp,  93, 5
\bibitem[Baum et al. 1988]{Ba88}
Baum, S. A., Heckman, T. M., Bridle, A., Van Breugel, W.,
 Miley, G. K., 1988, \apjs, 68,  643
\bibitem[Baum and Heckman 1989a]{Baa89} Baum, S. A., and 
Heckman, T., 1989, \apj,  336, 681
\bibitem[Baum and Heckman 1989b]{Bab89} Baum, S. A., and 
Heckman, T., 1989, \apj,  336, 702
\bibitem[Burrows  et al. 1991]{bur91} 
Burrows C.~J.~et al.~1991, \apjl,  369, L21
\bibitem[Carswell 1984]{Ca84} Carswell, R.F., Bauldwin, J. A., 
Atwood, B. and Phillips, M. M., 1984, \apj  286, 464
\bibitem[Christiansen  et al. 1977]{Ch77} 
Christiansen, W.\/N.\/, 
 et al., 1977, \mnras,  181, 183 
\bibitem[Clark  et al. 1997]{Cl97} Clark  et al., 1997,
\mnras,  286, 558
\bibitem[Cohen 1983]{Co83} Cohen, R.\/ D.,  1983, \apj,  273, 489.
\bibitem[Conway  et al. 1990]{Co90} Conway J.\/E., 
Cornwell T.\/J.\/, Wilkinson P.\/N.\/, 1990, \mnras,  246, 490
\bibitem[Dopita and Southerland 1995]{Do95}Dopita, M., A., and
Southerland, R., S., 1995, \apj,  455, 468
\bibitem[Evans 1991]{Ev91} Evans, Ian, private communication.
\bibitem[Ferguson  et al. 1997]{Fe97}Ferguson, J., W., 
Korista K., T., Baldwin J., A., and 
Ferland G., J. 1997, \apj,  287, 122
\bibitem[Filippenko 1985]{Fp85} Filippenko, A.\/, 1985, \apj,  289, 475 
\bibitem[Fraix--Burnet  et al. 1991]{FB91} Fraix-Burnet, 
D.\/, Golombek, D.\/, Macchetto, F.\/, Nieto, J.\/,  Lelievre, G.\/, 
Perryman, M.\/ A.\/ C.\/, Di Serego Alighieri, S.\/ 1991, \aj, 101,  88.
\bibitem[Grimberg et al. 1999]{Gr99} Grimberg, B. I., Simkin, S.\/ M., 
and Sadler, E. \/M., 1999, \apj, submitted (MS\#39792)
\bibitem[Halpern  et al. 1996]{Ha96} Halpern, J.\/ P., Eracleous, 
M.\/, Filippenko, A.\/ V., Chen, K.\/ , 1996 \apj,  464, 704
\bibitem[Hanish and White 1994]{Ha94}Hanish, R.\/J.,\/ and
White.\/R.\/L., ``The Restoration of HST Images and Spectra,'' 1994,
Space Science Telescope Institute, Baltimore, MD
\bibitem[Hansen  et al. 1987]{Ha87} Hansen, L.\/, 
Jorgensen, H.\/ E.\/, Norgaard-Nielsen, H.\/ U.\/, 1987, \aaps, 
 71,  465
\bibitem[Heckman  et al. 1985]{He85} Heckman, T. M., Illingworth,
G. D., Miley, G. K., Van Breugel, W. J. M.  \apj, 1985,  299, 41
\bibitem[Jenkner  et al. 1990]{Je90}Jenkner, H., 
Lasker, B. M., Sturch, C.R., McLean, B.J., Shara, M. M. and 
Russell, J. L., 1990, \aj, 99, 2081 
\bibitem[Knapp  et al.  1990]{Kn90}Knapp, G. R., Bies, W. E., 
Van Gorkom, J. H., \aj, 1990,  99,  476
\bibitem[Krist 1994]{Kr94} Krist, J.\/ E., HST Tiny Tim 
Software, Version 4.0b release  September 1994 
\bibitem[Lasker  et al. 1990]{La90}Lasker, B. M., 
Sturch, C. S., McLean, B. J., Russell, J. L., Jenkner, H., 
and Shara, M. M. 1990, \aj, 99, 1019
\bibitem[Mackenty  et al. 1994]{Mk94}  MacKenty , J.\/W.\/, 
Simkin, S.\/M.\/, Griffith, R.\/E.\/, Ulvestad, J.\/S.\/, 
and Wilson, A.\/S.\/  1994, \apj, 435, 71
\bibitem[Morgan 1995]{Mo95}Morgan, J. A. 1995, in  ``Astronomical
Data Analysis Software and Systems IV'', PASP Conf Series 77, editors R.
A. Shaw, H. E.  Payne, and J. J. E. Hayes, p.~129.
\bibitem[Osterbrock 1989]{Os89} Osterbrock D.\/,``Astrophysics of 
Gaseous Nebulae and Active Galactic Nuclei, 1989, page 134, 
University Science Books, Mill Valley, CA.
\bibitem[Perley  et al. 1997]{Pr97} Perley,R.\/A., 
R\"{o}ser, H.-J.\/, and Meisenheimer, K.\/ 1997,  \aap,  328, 12
\bibitem[Prieto et al. 1993]{Pr93} Prieto, M. A., Walsh, J. R.,
Fosbury, R. A. E., Di Serego Alighieri, S. 1993, \mnras, 263, 10
\bibitem[Reynolds 1994]{Re94} Reynolds J.\/E., 1994, ATNF 
Technical Document Series 39.3/040
\bibitem[Meisenheimer and  R\"{o}ser 1987]{MR87}  
R\"{o}ser, H.-J.\/, and Meisenheimer, K.\/ 1997,  \apj ,  314, 70
\bibitem[Robinson 1987]{Ro87}Robinson, A., Binette, L., Fosbury, R.
A.,E., and Tadhunter, C. N., 1987, \mnras,  227, 97
\bibitem[Russell  et al. 1990]{Ru90}Russell, J.L., 
Lasker, B. L., McLean, B. J., Sturch, C. R., and Jenkner, H., 
1990, AJ, 99, 2059 
\bibitem[Sault 1994]{Sa94} Sault R.\/J., Wieringa M.\/H.\/, 
1994, \aaps,  108, 585
\bibitem[Sault 1995]{Sa95} Sault R.\/J., Teuben P.\/J, 
Wright M.\/C.\/H, \/1995, eds. R. Shaw, H.\/E.\/ Payne,
J.\/J.\/E.\/ Hayes,  Astronomical Data Analysis Software and 
Systems IV, A.\/S.\/P.\/,
 77, 433
\bibitem[Simkin 1979]{Si79} Simkin, S.\/M.\/, , 1979, \apj,  234, 56 
\bibitem[Storchi-Bergmann   et al. 1996)]{SB96}Storchi-Bergmann, T.,
Wilson, A., S., Mulchaey, J., S., and Binette, L., 1996, \aap,  312,
357
\bibitem[Su and Simkin 1980]{Su80}Su,\/ H-J., and Simkin,\/ S.\/ M., 1980, \apjl,  238, L1 
\bibitem[Sulentic  et al. 1995]{Su95} Sulentic, J. W., 
Marziani, P., Zwitter, T., Calvani, M.  1995, \apjl,   438,  L1
\bibitem[Tadhunter  et al. 1989]{Ta89} Tadhunter, C. N., 
Fosbury, R. A. E., and Quinn, P. J., 1989, \mnras,  240, 225
\bibitem[Thomson  et al. 1995]{Th95} Thomson, R.\/ C.\/, 
Crane, P.\/, and Mackay, C.\/D.\/, 1995, \apj,  446, L93
\bibitem[Tingay  et al. 1999]{Ti99}Tingay, S.\/J.,  et al., 1999,
submitted to the \apjlett
\bibitem[Veilleux and  Osterbrock 1987]{VO87}Veilleux, S., and 
Osterbrock, D., \aaps, 1987,  63, 295
\bibitem[Viegas and  de Gouveia Dal Pino 1992]{Vi92}Viegas,\/S.\/M.\/
de Gouveia Dal Pino,\/E.\/M.,\/ 1992, \apj,  384, 467
\bibitem[Villar-Martin  et al. 1998]{Vi98} Villar-Martin, M.,
 et al. 1998, \aap,  332, 479
\bibitem[Weir 1991a]{wea91}  
Weir, N.~ 1991a,  ESO Workshop No. 38, P.~J.~Grosbol and R.~H.~Warmels, 
ESO: 115
\bibitem[Weir 1991b]{web91} 
Weir, N.~ 1991b,  10th International Workshop on Maximum Entropy 
and  Bayesian Methods, editors W.~T.~Grandy and L.~H.~Schick, 
(Dordrecht: Kluwer), 275
\bibitem[Westphal et al.~ 1982]{wes82} Westphal et al.~1982, The Space
Telescope Observatory, D.~N.~B.~Hall, NASA: CP-2244, 28
\bibitem[WFPC Handbook, 1989]{WF89}  Hubble Space Telescope Wide 
Field and Planetary Camera Instrument handbook, version 2.0, May, 1989
\bibitem[Wilson and Baldwin 1985]{wil85}  
Wilson A.~S.~ and Baldwin J.~A.~ 1985, \apj, 289, 124
\hspace*{-12.0in}

\end{thebibliography}
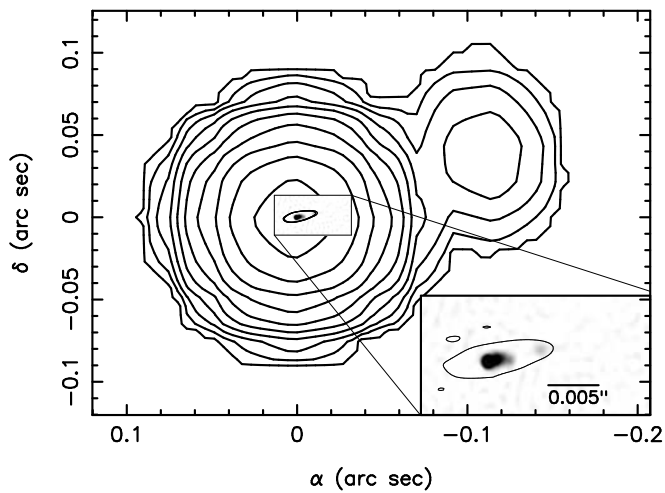
\end{document}